\documentclass[slac_one]{revtex4}
\usepackage[latin1]{inputenc}
\usepackage[brazil,english]{babel}
\usepackage{graphicx}
\usepackage{fancyhdr}
\begin{document}

\title{Cosmography with Galaxy Clusters} 

%

\author{M. Soares-Santos}
\email{marcelle@fnal.gov}
\affiliation{Fermi National Accelerator Laboratory, Batavia, IL 60510, USA}
\affiliation{\foreignlanguage{brazil}{Universidade de São Paulo, 
 São Paulo, SP 05508-090, Brazil}}

\author{R. R. de Carvalho}
\affiliation{VST Center at Naples, Napoli, 80131, Italy}
\affiliation{\foreignlanguage{brazil}{Instituto Nacional de Pesquisas 
Espaciais, São José dos Campos, SP 12227-010, Brazil}}

\author{F. La Barbera}
\affiliation{Osservatorio Astronomico di Capodimonte, Napoli, 80131, Italy}

\author{P. A. A. Lopes}
\affiliation{\foreignlanguage{brazil}{Universidade do Vale do Paraíba, São José
dos Campos, SP 12244-000, Brazil}}

\author{J. Annis}
\affiliation{Fermi National Accelerator Laboratory, Batavia, IL 60510, USA}

\begin{abstract}
In the present work we focus on future experiments using
cluster abundance observations to constraint the Dark Energy equation
of state parameter, w. To obtain tight constraints from this kind of
experiment, a reliable sample of galaxy clusters must be obtained
from deep and wide-field images. We therefore present the computational
environment (2DPHOT) that allow us to build the galaxy catalog from the
images and the Voronoi Tessellation cluster finding algorithm that we
use to identify the galaxy clusters on those catalogs. To test our
pipeline with data similar in quality to what will be gathered by
future wide field surveys, we process images from the Deep fields
obtained as part of the LEGACY Survey (four fields of one square degree
each, in five bands, with depth up to r'=25).
We test our cluster finder by determining the completeness and purity
of the finder when applied to mock galaxy catalogs made for the
Dark Energy Survey cluster finder comparison project by Risa
Wechsler and Michael Busha. This procedure aims to understand the
selection function of the underlying dark matter halos.
\end{abstract}

\maketitle

\thispagestyle{fancy}


\section{INTRODUCTION} 
Observational evidences such as the luminosity-redshift relation of type Ia 
supernovae, 
the power spectrum of the cosmic 
microwave background radiation 
and the distribution of large scale 
structures in
the Universe indicate that the total energy density of the Universe 
is dominated by the so-called dark energy, a slowly varying fluid with negative
pressure. The nature of dark energy is an important open question connected
to the fundamental theories of gravity  and therefore the scientific 
community is investing great efforts in developing tools to address this 
problem. 
The best fit to current observations places Einstein's cosmological constant in
the role of Dark Energy (the $\Lambda$-CDM model),  but the constraints on the 
equation of state parameter ($w \equiv p/\rho$, where $w=-1$ is the signature 
of the cosmological constant) are not definitive.
The combination of all currently available  datasets allowed to constraint  
$w$  up to  $10-20\%$ uncertainty and
the next generation of experiments aim to improve it to $\lesssim 1\%$. 
\cite{Albrecht:06} 

One of the techniques used to study Dark Energy observationally is based on 
galaxy cluster surveys \cite{Haiman:01,Evrard:02}.
The cluster redshift distribution ($d^2 N/dzd\Omega$) is the number of
clusters per unit redshift per unit solid angle, which  is obtained as the 
product of the volume element of the survey  ($d^2V/dzd\Omega$) and the number 
density of clusters $n(z)$. Both quantities are sensitive to the cosmological 
model: the former through the angular diameter distance $d_A(z)$ and the
Hubble parameter $H(z)$;
while the later, being the integral of the halo mass function $f(M,z)$, 
depends on the linear growth of density perturbations.
The evolution of $f(M,z)$ in the context of the linear growth theory,
is studied using N-body simulations \cite{Evrard:02}
and it is shown that the number of clusters at redshifts larger than $0.5$
decreases strongly with $w$ and that this effect is more evident at the 
high mass tail
of the distribution. 
Therefore, to improve the constraints on $w$ using galaxy clusters, it is
necessary to obtain a 
well-understood selection of a large sample of massive high redshift clusters.
However, the abundance of such objects is as low as a few ones per square 
degree 
and this implies that the survey must have a large area coverage.
A survey covering a few thousands square degrees up to 
$z \simeq 1.4$ would lead to less than $5\%$ uncertainty on $w$, if the 
cluster masses are determined with uncertainty  $<10\%$.
\cite{Levine:02,Lima:04}.
This means that galaxy clusters surveys alone can provide a significant 
improvement in the current status of the Dark Energy investigations and 
the combination with data obtained from other techniques will allow to 
achieve the subpercent 
uncertainty level.  


The accomplishment of a dark energy study using galaxy clusters as 
described above requires the development of improved image processing tools
and cluster finding algorithms. The image processing must perform accurate 
star-galaxy separation at faint flux levels 
(up to magnitude $r' \sim 25$, two magnitudes fainter than the
current wide field surveys) and produce a galaxy catalog with high 
completeness and low contamination. The cluster finder  will  
take this galaxy catalog and produce a cluster catalog. The leading 
algorithms currently in use \cite{Koester:07}
are very efficient up to $z \sim 0.5$, but
a significant improvement is needed to achieve the same efficiency at
the redshift range of 
interest for dark energy studies. 
Here we present results for both the 2DPHOT image processing
package \cite{LaBarbera:08}
and the Voronoi Tessellation (VT) cluster finder algorithm \cite{Ramella:01}. 
To test our
pipeline with data similar in quality to what will be gathered by
the next generation of 
wide field surveys, we process images from the Deep fields
obtained as part of the LEGACY Survey \cite{Olsen:08}.
To understanding the selection function of the dark matter halos 
we compute the completeness and purity 
of the  cluster catalog using mock galaxy catalogs made by Risa Weschler
and Michael Busha for the Dark Energy Survey cluster finders comparison 
project. The present setup
of our cluster finder produces a catalog with high completeness, but
purity needs to be improved.

\section{2DPHOT}
\begin{figure}[ht]
\includegraphics[width=50mm]{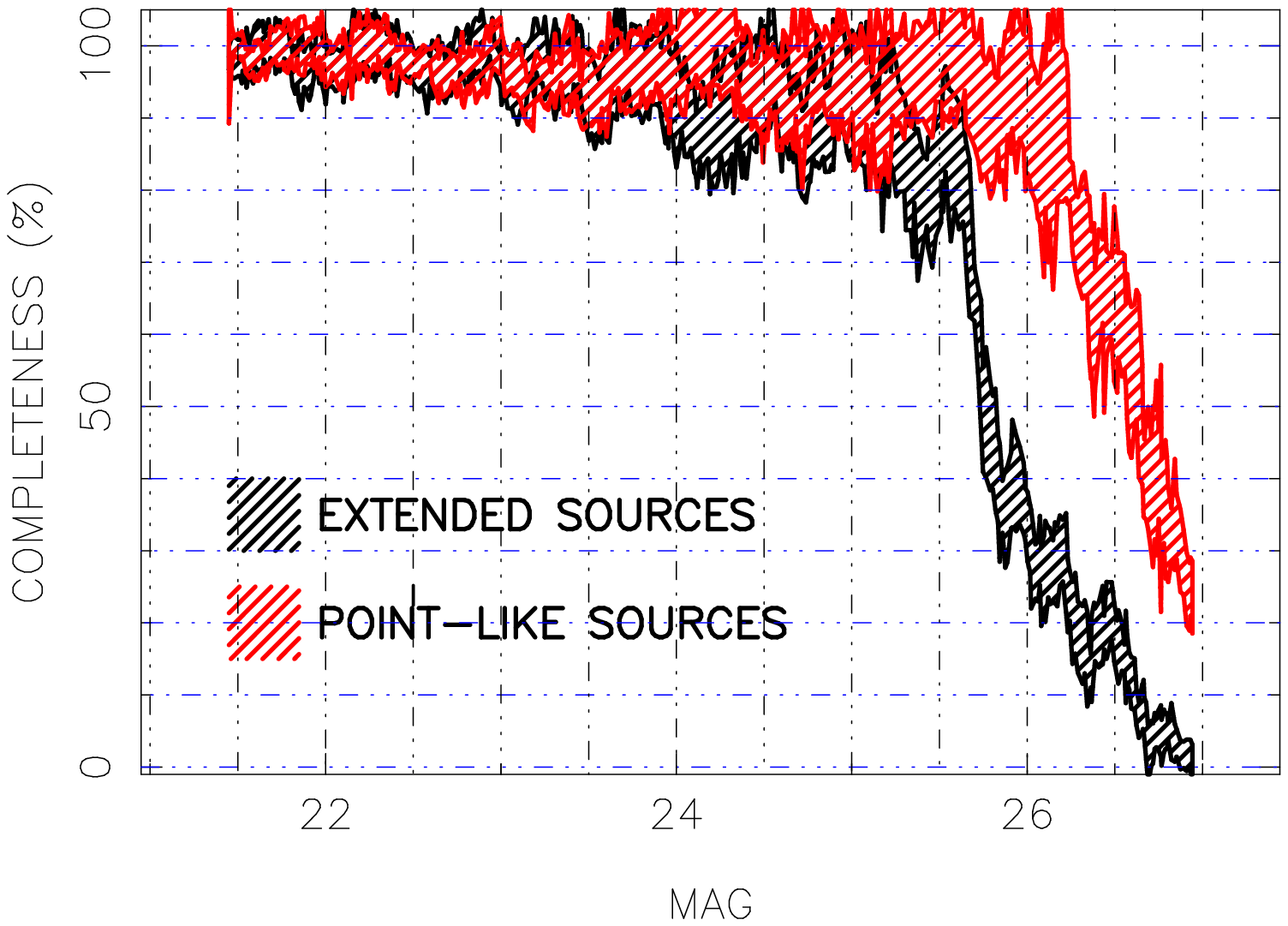} \hspace{10mm}
\includegraphics[width=50mm]{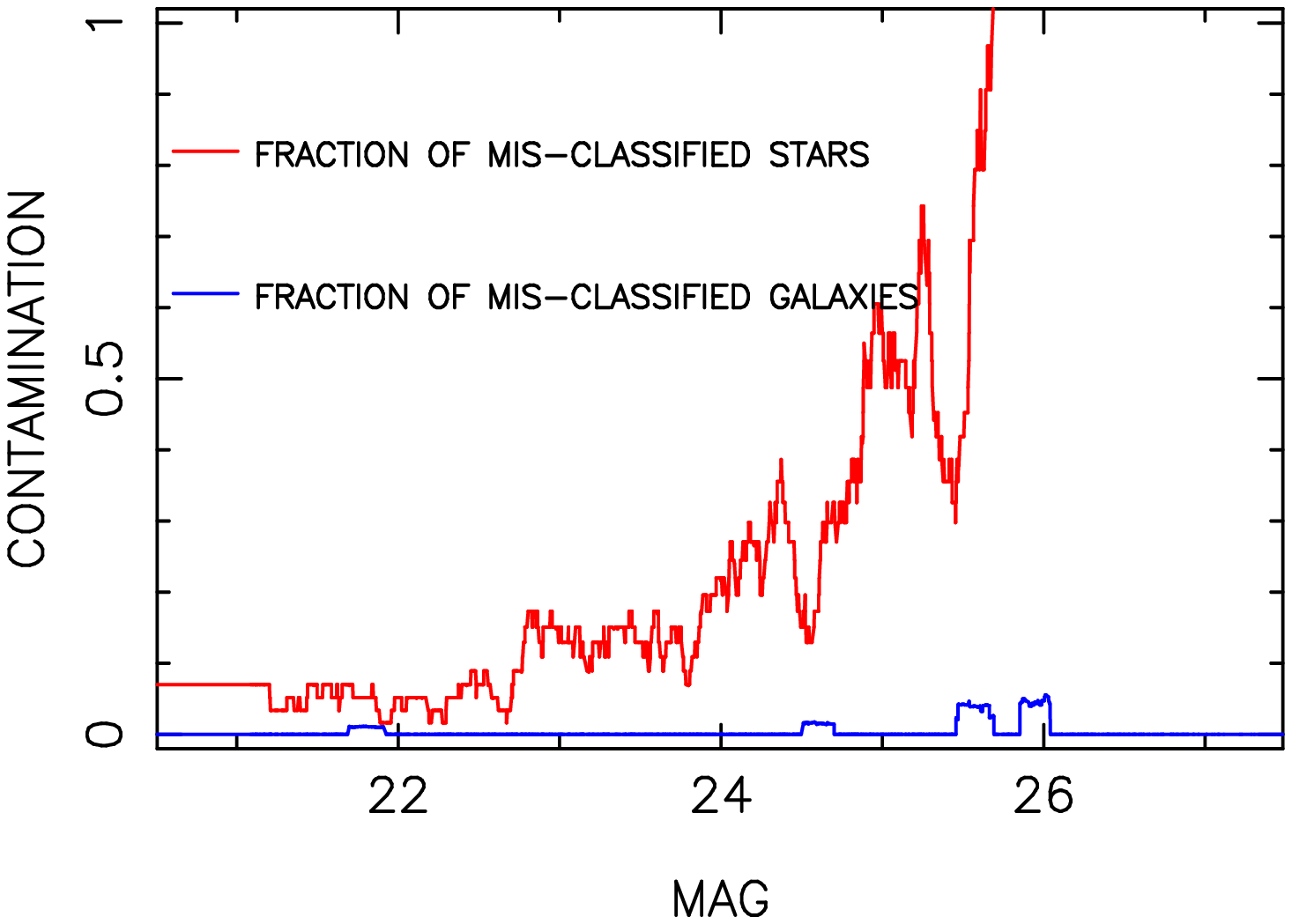}
\caption{Completeness and contamination as function of the
$r'$  magnitudes,  for one subimage of the 
Deep LEGACY survey.} \label{2dphotresults}
\end{figure}

2DPHOT
was developed to perform detection, global
photometric measurements and surface photometry of objects in a given 
image \cite{LaBarbera:08}.  
In order to estimate the completeness of the galaxy catalog, 2DPHOT adds
simulated galaxies and stars to the image, 
using the point spread function and the
distribution of galaxy structural parameters 
obtained from the processed image. Coordinates and magnitudes are 
randomly assigned 
within the range of observed objects. 
The image is reprocessed and
the fractions of recovered
and misclassified artificial
objects correspond to the completeness and contamination of the processed 
catalog, respectively. 
We process the Deep fields of the LEGACY survey, which  is not wide
enough to set constraints on $w$, but has the same depth and image quality 
(seeing $\sim 0.7''$) of future wide surveys, being the ideal dataset 
to calibrate our pipeline. 
The results of this processing (Fig. \ref{2dphotresults}) shows  
that the galaxy catalog has negligible contamination and 
completeness $>90\%$ up to $r'=24.5$. After processing the images in the
remaining bands, we have, for each of the  Deep fields, 
a galaxy catalog to be used as input for our cluster finder.  

\section{VORONOI TESSELLATION CLUSTER FINDER}

The VT algorithm analyzes the spatial distribution
of galaxies in 2 dimensions \cite{Ramella:01}. To account for the depth of the
catalog, we work in shells of either photometric redshift or magnitude.
The magnitude binning is particularly useful when there is no photometric 
redshift available for a given dataset. 
In each shell, a unique partitioning of the plan into 
convex cells, each containing a single galaxy, is obtained.  
The local
density is then measured as the inverse of the area of each cell and 
the clusters of galaxies are detected as local density enhancements with 
respect to the density distribution
of background galaxies. VT is a non-parametric
method that requires no smoothing of the input data 
and allows overdensities
such as groups and clusters of galaxies to be identified independently of 
their 
geometry. 
To detect the clusters, we 
compare the distribution of cell areas with the distribution expected from 
a Poissonian process. The deviation 
between these distributions is  negligible in the regions far from the 
cluster and gets larger as we approach an overdensity.  We then set a 
density threshold below which we will consider the cells as being part of 
a cluster candidate. This threshold is the point where the 
cumulative distribution of cells becomes stiffer than the Poissonian
distribution and corresponds to $\sim 80\%$ of the distribution.     
We then estimate the probability 
that each cluster candidate has to have been 
generated by a background (Poissonian) fluctuation and we reject the ones 
for which this probability is larger than $5\%$. We finally use a percolation
scheme to combine the cluster candidates in each bin 
into a unique list of clusters.   
We apply the VT on mock galaxy catalogs to estimate the completeness and 
purity of the resulting catalogs. Mock catalogs are produced using 
$\Lambda$CDM N-body simulations to generate the dark matter halos which 
are then populated with
galaxies following the luminosity function observed in the local Universe.
The mock catalogs are as deep as $z \sim 1.4$ and contain magnitudes,
redshift and photometric redshift for each galaxy.
In order to quantify the completeness and purity, we used a cylindrical 
matching scheme in which each VT cluster is matched to a mock halo within 
a cylinder with height $\Delta z = 0.05$ and 1.5Mpc radius. 
Completeness is defined as the fraction of matched halos
while purity is the fraction of matched VT clusters. 
Using magnitude binning we obtained completeness $>95\%$, but purity was 
limited to $60\%$. The analysis using photometric redshift is still in 
progress, but we expect to improve purity to $\sim 90\%$ using an 
optimized binning and percolation setup. 

\section{CONCLUSIONS}
In this work we show preliminary results about the development of tools
to perform galaxy clusters detection from deep and wide field images,
aiming at future dark energy studies using galaxy cluster survey data. 
We take the Deep LEGACY survey data, which 
covers a small area  (4 sq deg) but is similar in quality to what will
be gathered by these future experiments, as a benchmark data set to
test our pipeline. 
We use the 2DPHOT package to process the Deep images and obtain the galaxy 
catalogs with $90\%$ completeness up to $r' = 24.5$. 
The VT cluster finder is being calibrated using mock galaxy catalogs.
The current results show high completeness up to redshift $1.4$, but 
purity has to be improved. After this calibration process, we intend to apply 
the VT cluster finder to the Deep catalogs produced using 2DPHOT to test
our full pipeline on real high redshift data for the first time.      

\begin{acknowledgments}
Work supported by the Brazilian National Council of Research (CNPq).
\end{acknowledgments}


\end{document}